\documentclass{article}

\usepackage{graphicx}
\usepackage{latexsym}

\begin{document}

\title{Percolation on Growing Lattices}

\author{Daniel Tiggemann\\
Institute of Theoretical Physics, University of Cologne\\
50937 K\"oln, Germany, European Union\\
dt@thp.uni-koeln.de}

\maketitle

\begin{abstract}
In order to inverstigate the dependence on lattice size of several observables
in percolation, the Hoshen-Kopelman algorithm was modified so that growing
lattices could be simulated. By this way, when simulating a lattice of size
$L$, lattices of smaller sizes can be simulated in the same run for free,
saving computing time.

Here, site percolation in three dimensions was studied. Lattices of up to
$L=5000$, with many $L$-steps in between, have been simulated, for 
occupation probabilities
of $p=0.25$, $p=0.3$, $p=p_c=0.311608$, and $p=0.35$.
\end{abstract}

\section{Introduction}

Percolation is a thoroughly studied model in statistical physics.\cite{1}  
The algorithm invented by Hoshen and Kopelman in 1976 allows for examing 
large percolation lattices using Monte Carlo methods,\cite{2} as for simulating
a $L^d$ lattice only a hyperplane of $L^{d-1}$ sites has to be stored.
The normal way to do
a simulation using the HK algorithm is to choose an $L$ and an occupation
probability $p$, and then walk through the lattice in a linear fashion, 
calculating interesting observables (like cluster numbers) on the fly or at
the end. When one is interested in the dependance of these observables on
either $p$ or $L$, one has to repeat these simulations with varying values
of these parameters.

In 2000 Newman and Ziff published an algorithm which allows to simulate
percolation for all $0 \le p \le 1$ in one run, making it very easy to
study percolation properties in dependance on $p$.\cite{3} Unfortunately,
their algorithm lacks a desirable property of the HK algorithm: They need
to store the full lattice of $L^d$ sites, whereas for HK only memory for
$L^{d-1}$ sites is needed.

A slight modification of the HK algorithm allows to simulate percolation
for many $1 \le L \le L_{\mathrm{max}}$ in one run, with only a small
performance penalty (depending on the number of intermediate $L$ which are
chosen for investigation). The biggest advantage of the HK algorithm, 
small memory consumption, is preserved. The modified algorithm needs to
store three times as much sites as the original one (a constant factor),
but this is still $\propto L^{d-1}$, meaning big advantages for low dimensions.

This modified algorithm shall be presented in this paper, along with
simulation results for site percolation on the cubic lattice, i.~e.~$d=3$.

\section{Computational method}

The traditional HK algorithm works by dividing the $L^d$ lattice into $L$
hyperplanes of $L^{d-1}$ sites, then investigating one hyperplane after the
other. One advantage of this simple scheme is that it allows for 
domain decomposition and thus 
parallelization.\cite{4}

Another approach is to work recursively on growing lattices: first simulate
a lattice of size $L^d$, then advance to $(L+1)^d$ by adding a shell with 
a thickness of one site onto the old lattice. This way, when simulating
an $L_{\mathrm{max}}^d$ lattice, all sub-lattices of sizes $L<L_{\mathrm{max}}$
can be investigated; the investigation of intermediate lattices may
take some additional time (for counting clusters etc.), but the
time spent on generating the full lattice minus time for investigation
is the same, because the same number of
sites is generated as with the traditional approach.

\subsection{A modified Hoshen-Kopelman algorithm}

\begin{figure}[htbp]
\centerline{\includegraphics[width=12cm]{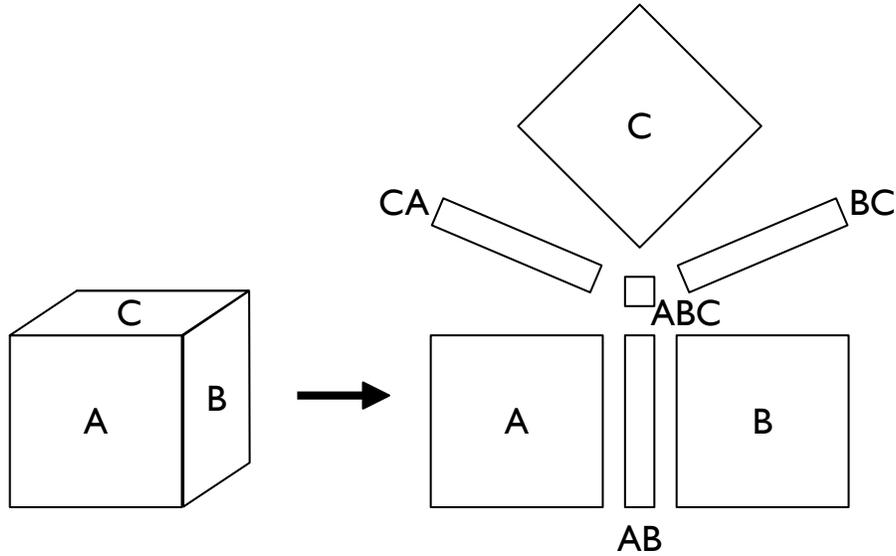}}
\vspace*{8pt}
\caption{Decomposition of the surface of the cubic lattice. Rear faces
X, Y, and Z are not shown. Due to the
Hoshen-Kopelman algorithm, in $d$ dimensions, only a $d-1$ hyperplane needs to
be stored, here for the growing cube only its surface. When going from an
$L$-cube to an $L+1$-cube, the surface grows. As edges and corners have to be
treated specially, they are stored separately. For a $L \times L \times L$ cube,
the faces (A, B, C) are $L-1 \times L-1$, the edges (AB, BC, CA)
$L-1 \times 1$, and the corner (ABC) is $1 \times 1$.
\label{f1}}
\end{figure}

This recursive approach was chosen for this paper. It works as follows:
Imagine a cube of size $L^3$. It has six faces, labelled A, B, C, X, Y, Z
(cf.~fig.~\ref{f1}). In order to increase the size $L$ of the cube by one, 
three new faces A', B', C' with a thickness of one are slapped onto the
old faces A, B, C; additionally, edges AB, BC, and CA, and a corner ABC
are added, forming a full $(L+1)^3$ cube.

\begin{figure}[htbp]
\centerline{\includegraphics[width=12cm]{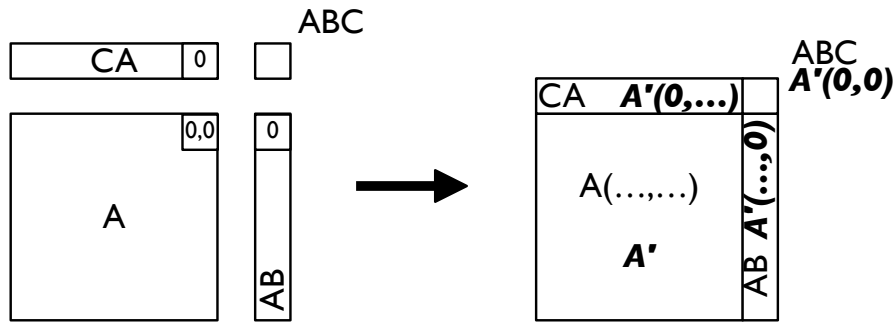}}
\vspace*{8pt}
\caption{When growing the cube from $L$ to $L+1$, the surface which needs to
be stored for the Hoshen-Kopelman algorithm, needs to grow, too. The faces of
the cube, here A, grow from $L-1 \times L-1$ (A, plain font in figure)
to $L \times L$ (A', bold italic in figure).
For the largest part of A', A'$(i>0,j>0)$, the back neighbors are sites of
A, but for the inner rim of A', A'$(i=0,j>0)$ resp.~A'$(i>0,j=0)$, the
back neighbors are CA resp.~AB. For A'$(0,0)$, it is ABC.
\label{f2}}
\end{figure}

Using the HK algorithm (both the original and the modified version), each
newly generated site has $d$ old neighbors. For $d=3$, we have top, left, and
back neighbors. For the sake of simplicity, we assume here that top and left
neighbors are within a new face A' (resp. B', C'), while back neighbors are
in the old face A. Due to geometry, site numbering is shifted by one: the site
A'$(i>0,j>0)$ has a back neighbor of A$(i-1,j-1)$.
A'$(0,j>0)$ has a back neighbor CA$(j-1)$, A'$(i>0,0)$ has a back neighbor
AB$(i-1)$, and A'$(0,0)$ has ABC. Thus edges and the corner need to be
treated specially (cf.~fig.~\ref{f2} for more clarity).

The site A'$(i,j)$ has the left neighbor A'$(i+1,j)$ and the top neighbor
A'$(i,j+1)$, when we start working inwards from A'$(L,L)$.
This way, we do not need to store
A and A', but can overwrite A with the values of A'. The same trick is used
in the original HK algorithm for storing only one hyperplane instead of two.

This way we need three faces of size $(L-1)^2$, three edges of $L-1$ and
one corner of size 1, for a total of $3 L^2 - 3 L + 1$ sites, in order to
simulate a system of size $L^3$. For the traditional approach, we need
$L^2$ sites, meaning one third, a fixed factor. But we need an additional
label array for
both methods, so that memory consumption does not differ drastically.

As the new approach has about the same speed and only moderately higher
memory consumption than the traditional one, it is a viable alternative.

One important thing to keep in mind is that for a single run, results for
$L_1$ and $L_2 > L_1$ are {\it not} statistically independent, e. g. results
for $L$ and $L+1$ are naturally highly correlated. This problem can be
alleviated by averaging over many runs with different random numbers, because
results for different random numbers are suitably statistically independent,
as long as the used random number generator has a sufficiently high quality.

\subsection{Fully periodic boundary conditions}

For small lattices, open boundaries would mean a strong distortion of results,
in the most cases proportional to $1/L$ (surface divided by volume). In order
to get rid of these influences, it is necessary to use periodic boundary
conditions, i.~e. sites on an open border are connected to corresponding sites
on the opposite border.

This is also possible for the modified HK algorithm. It is necessary to store
not only the front faces A, B, C, but also the rear faces X, Y, Z. When we
choose to obtain observables for an intermediate lattice size $L$, we connect
sites on A to Z, on B to Y, and on C to X.

When during periodicization a cluster on one face connects to the same cluster
on the opposite face, that cluster spans the whole systems and wraps around,
and thus can be identified as the infinite cluster. When such a cluster is
detected, the system percolates. It is possible that more than one such cluster
forms; the number of different spanning clusters is counted as $n_\infty$.

As the label array is modified by 
periodicization, we need to save a copy first, because we need the original
array to continue the simulation for larger $L$ afterwards. After the
periodicization, the modified label array contains all cluster properties;
these can be easily extracted and written out. The modified array can be
discarded afterwards.

Using this method, it is possible to use fully periodic boundary conditions
for intermediate $L$, although the lattice grows afterwards. Periodic b.~c.
mean a performance penalty and a doubling of memory consumption (as six
faces instead of three need to be stored). But the same is also true of
the traditional HK algorithm for periodic b.~c., were two instead of one
hyperplane need to be stored, still giving a factor of three in
memory consumption.

\subsection{Recycling of labels}

When simulating large lattices, the label array is filled with labels, to which
sites in the lattice point. Due to the way with which the HK algorithm (both
original and modified version) works, lots of these labels are indirect ones,
which point to other labels. Other labels correspond to clusters that are 
hidden in the bulk of the lattice and no longer touch a surface. It is
possible to get rid of these labels and thus free up precious space
by using a method known as Nakinishi recycling.\cite{5}

In order to support fully periodic b.~c., for each recycling all labels
represented in A, B, C, X, Y, Z, AB, BC, CA, ABC, have to be taken into
account. The method is fully analogous to that used for the traditional HK
algorithm.

\section{Results}

All simulations were done using the R$(471,1586,6988,9689)$ pseudo-random
number generator.\cite{6} The value of $p_c$ is not known exactly for
site percolation on the cubic lattice. For this paper it was chosen as
$p_c = 0.311608$.\cite{7} Simulations were done for $p=0.25, 0.3, p_c, 0.35$,
in order to investigate behavior below, at, and above the critical threshold.
For each value of $p$, 1000 runs with different random numbers were made and
used for averaging. 
Total required CPU time was 10000 hours on 2.2 GHz Opteron
processors.

\subsection{Cluster size distribution}

For the number $N_s$ of clusters of size $s$ in a lattice, we expect a
distribution $N_s \propto s^{-\tau}$ right at $p_c$. To make analysis
easier, we look at $n_s$, the number of clusters with at least $s$ sites:

\begin{equation}
n_s = \sum^\infty_{s'=s} N_{s'} \simeq k s^{-\tau+1}
\end{equation}

\begin{figure}[htbp]
\centerline{\includegraphics[width=10cm]{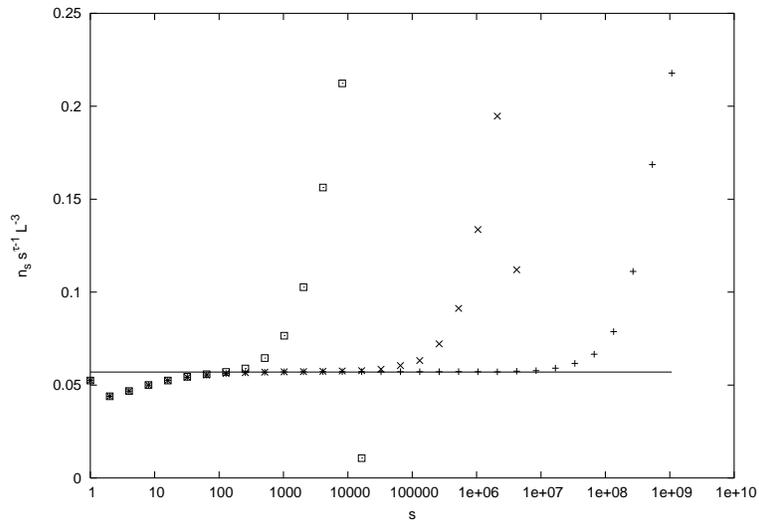}}
\vspace*{8pt}
\caption{Cluster size distribution $n_s$ for $L=50$ ($\Box$),
$L=500$ ($\times$), and $L=5000$ ($+$). By plotting $n_s s^{\tau -1} L^{-3}$,
all points should fall on the same line. For small $s$, we can see corrections
to scaling, for large $s$ the effects of the finite system size, even though
the boundary conditions are fully periodic. For larger $L$, the
finite size effects occur for larger $s$.
\label{f3}}
\end{figure}

In a log-log-plot, we would expect a straight line with slope $-\tau+1$.
Deviations from the power law would be hard to detect due to the logarithmic
scale, thus we plot $s^{\tau-1}n_s$ linearly on the $y$-axis. 
By varying $\tau$ until a flat plateau forms, we can estimate $\tau=2.189(1)$.
We can 
clearly see in fig.~\ref{f3} the deviation for small $s$, which are the same
for different $L$, and for large $s$, which differ for various $L$, as these
are caused by the finite system size.

For $p<p_c$, $n_s$ drops off sharply for small $s$, with $s$ slowly growing
for growing $L$ (not plotted). The same is true for $p>p_c$, but there does
an additional infinite cluster form (not plotted, too).

\subsection{Number of infinite clusters}

An infinite cluster is a cluster that spans the whole system. When using 
periodic b.~c., this cluster needs to wrap around, i.~e. span the whole system
and then periodically connect to itself.

According to theory, for $p<p_c$ no infinite cluster is expected, for $p=p_c$
about one with fractal properties, and for $p>p_c$ one with bulk properties. 
The critical threshold $p_c$ is well-defined only for infinite lattices,
where the number of infinite clusters (when these clusters are really
infinite) $n_\infty$ is exactly zero below $p_c$ and exactly one 
above $p_c$.
For finite lattices, it is
reasonable to define $p_c$ so that the average number of infinite clusters is
$1/2$, i.~e. a spanning cluster forms as often as it does not. Below $p_c$,
$n_\infty$ drops sharply to zero, above $p_c$ it goes up sharply to one. $p_c$
depends on the lattice size.

\begin{figure}[htbp]
\centerline{\includegraphics[width=10cm]{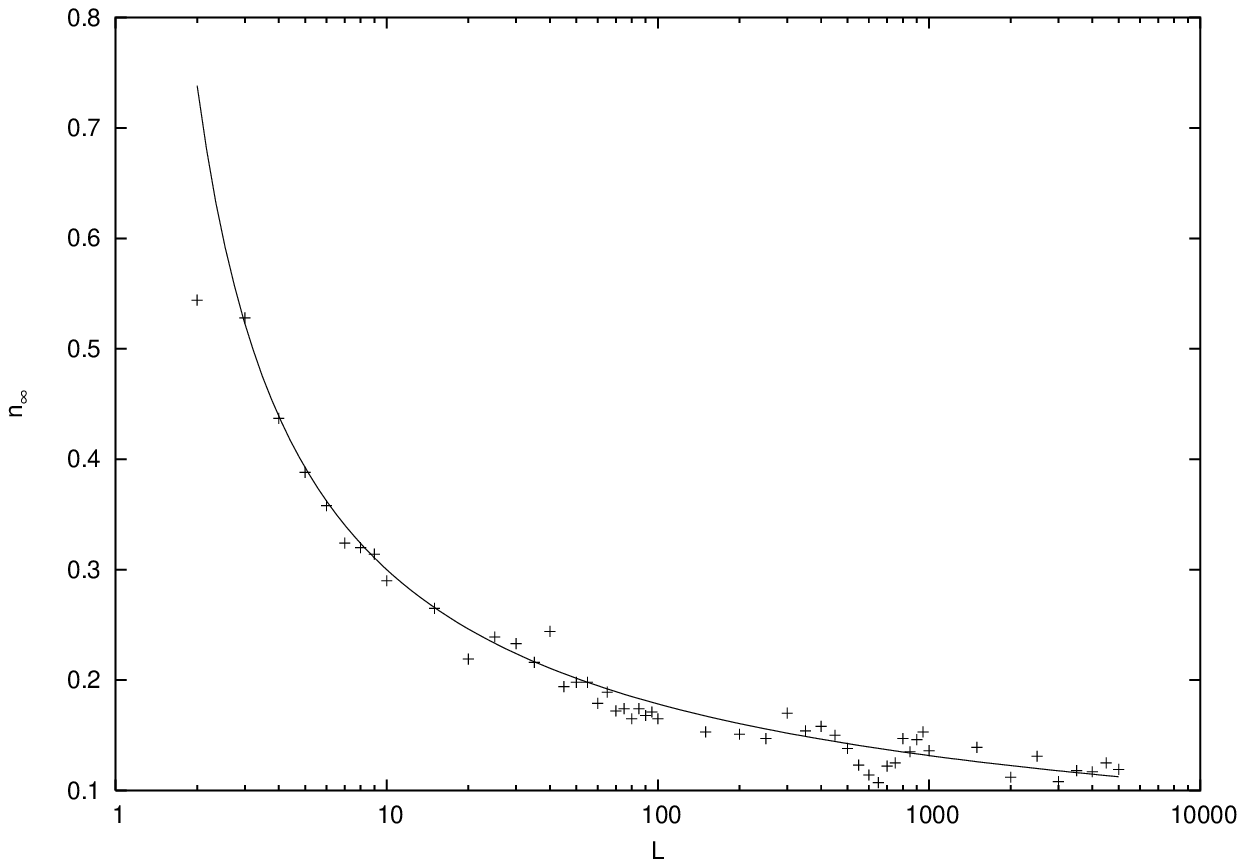}}
\vspace*{8pt}
\centerline{\includegraphics[width=10cm]{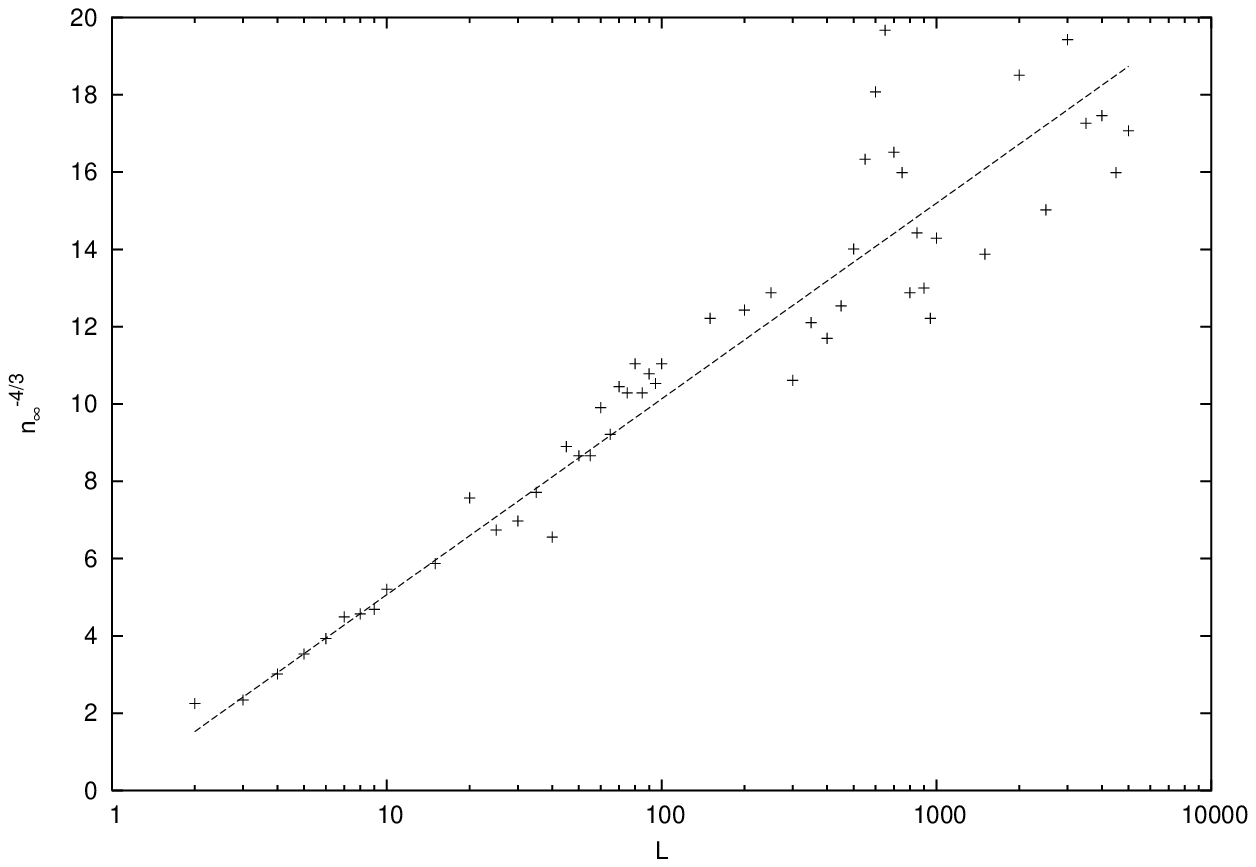}}
\vspace*{8pt}
\caption{Number $n_{\infty}$ of infinite clusters, at $p=0.311608 \simeq p_c$,
averaged over
1000 runs, depending on lattice size $L$. Right at $p_c$ we expect
$n_{\infty} = 1/2$, i.~e. a spanning cluster forms as often as it does not.
As here $n_{\infty}<1/2$, the value $p=0.311608$ chosen for this paper is
slightly to small; due to finite size effects, sufficiently large lattices
have to be simulated in order to see this, even when using periodic b.~c.
For the number of infinite clusters we get
$n_{\infty} = 0.3 (\log_{10}L)^{-3/4}$.
\label{f4}}
\end{figure}

For simulations at $p_c$, here a value of $p=0.311608$ was chosen. As can be 
seen form fig.~\ref{f4}, this value is slightly too small, but still very
near at the true value for $p_c$. An interesting behavior for $n_\infty$
results: $n_{\infty} \propto (\log L)^{-3/4}$.

\begin{figure}[htbp]
\centerline{\includegraphics[width=10cm]{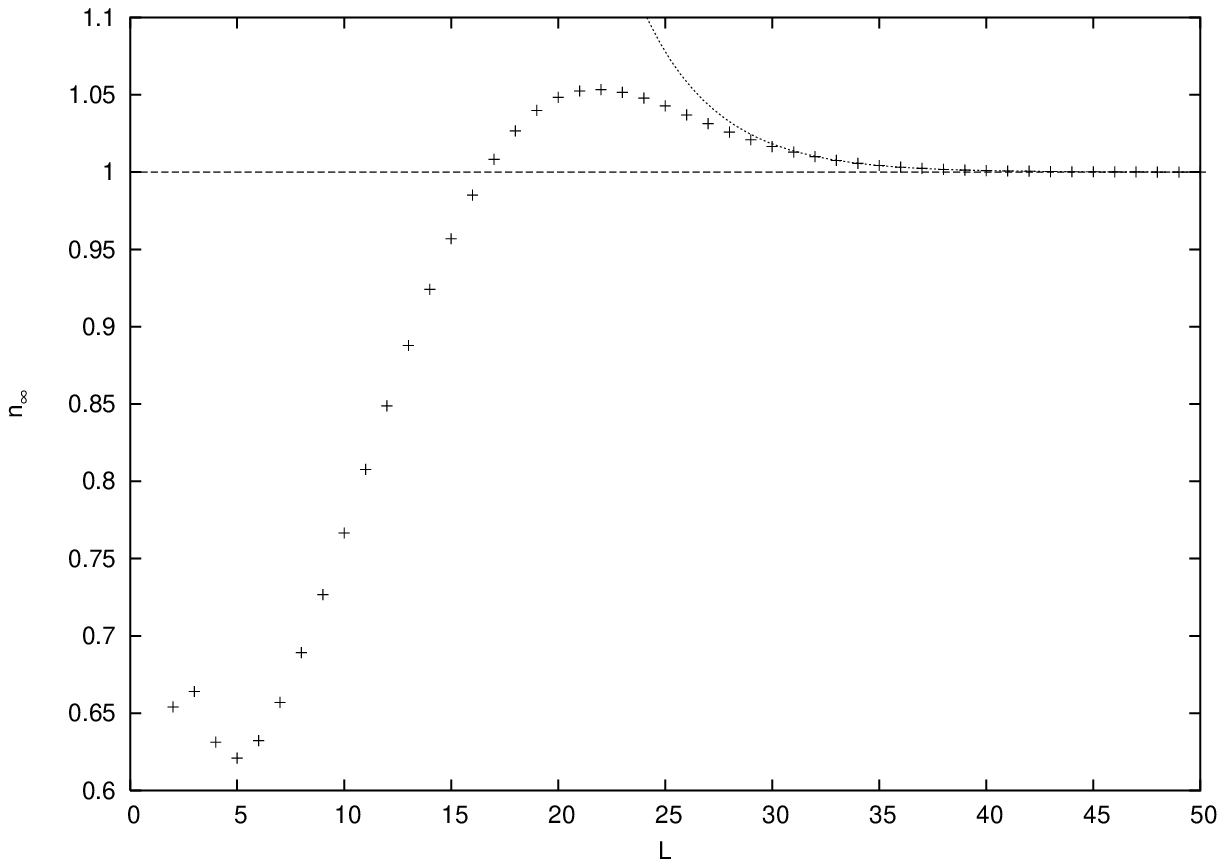}}
\vspace*{8pt}
\centerline{\includegraphics[width=10cm]{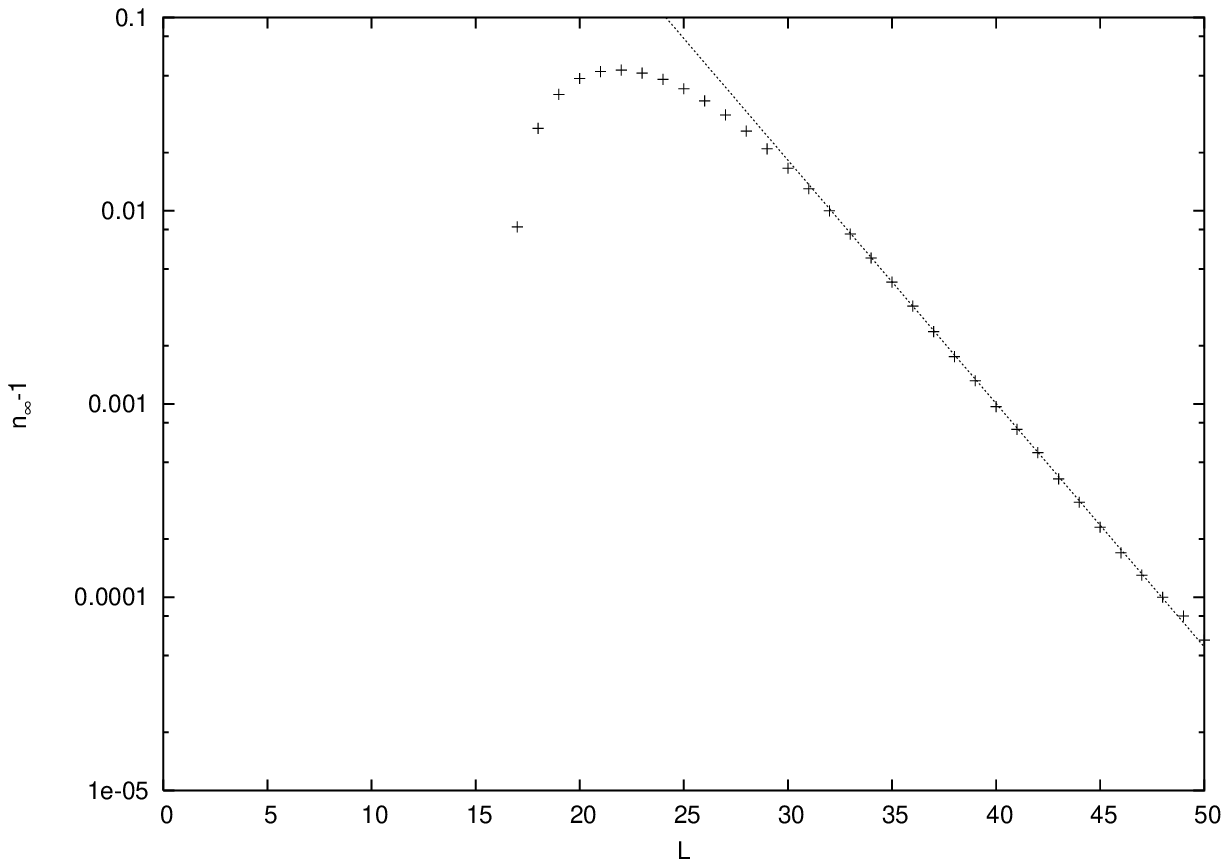}}
\vspace*{8pt}
\caption{Number $n_{\infty}$ of infinite clusters (i.~e. that wrap around
the system) for $p=0.35$, depending on lattice size. Theory predicts that only
one such cluster forms. However, for small lattice sizes a complex finite
size behavior can be seen: for $L<17$, less than one infinite cluster forms
on average, for $L>17$ more than one cluster forms, with a maximum at $L=22$,
afterwards $n_\infty$ drops to $1$ exponentially. The solid line corresponds to 
$1+e^{-0.29 (L-16.2)}$. For this plot, $10^7$ runs of up to $L=50$ have been
averaged.
\label{f7}}
\end{figure}

For $p>p_c$, $n_\infty$ is expected to be exactly equal one. This is true
for lattice sizes $L>50$, but for smaller lattices sizes, an interesting
behavior is visible (cf.~fig.~\ref{f7}): for very small lattices, less than
one infinite cluster forms on average, for slightly larger lattices, 
sometimes more than one such cluster forms.

For $p<p_c$, no infinite cluster forms for even small $L$. Only for very small
$L$, sometimes an infinite cluster appears. For larger $p$ this $L$ becomes
larger (not plotted).

\subsection{Size of largest cluster}

\begin{figure}[htbp]
\centerline{\includegraphics[width=10cm]{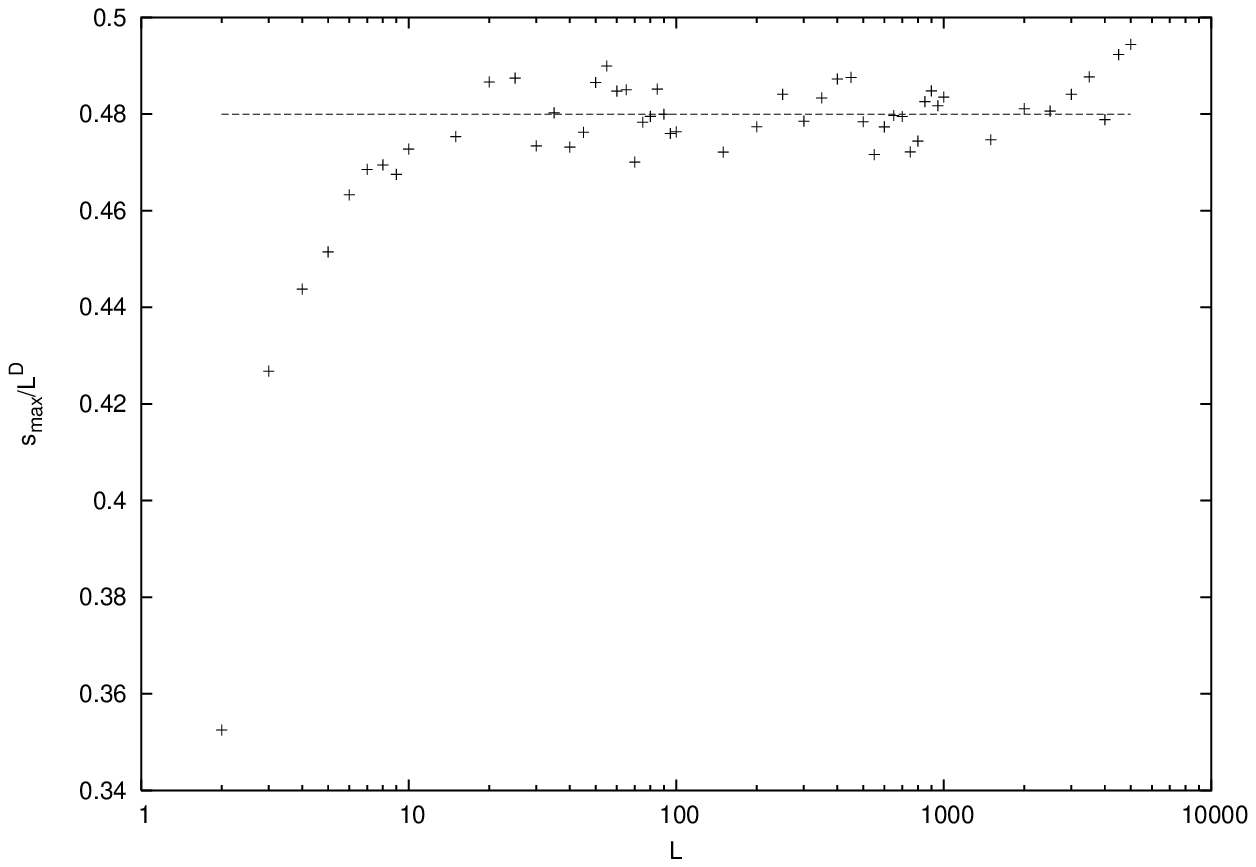}}
\vspace*{8pt}
\caption{Size $s_{\mathrm{max}}$
of the largest cluster (which is the infinite cluster,
if it forms), at $p=p_c$, scaled by $L^D$,
depending on simulated lattice size.
The cluster
grows $\propto L^D$, with a fractal dimension of $D=2.52(1)$. Thus, by scaling,
we get a straight line (the line visible in the plot serves as guide to the
eye). For very small lattice sizes $L$, corrections to scaling can be seen.
\label{f5}}
\end{figure}

As mentioned above, the infinite cluster has fractal properties at $p=p_c$,
meaning that it grows $\propto L^D$, with $D$ a non-integer value. From 
fig.~\ref{f5} we can extract a value of $D=2.52(1)$.

For $p>p_c$, the infinite cluster obtains bulk properties, i.~e.~it grows
$\propto L^D$ with $D=3$; it has no longer fractal properties (not plotted).

\begin{figure}[htbp]
\centerline{\includegraphics[width=10cm]{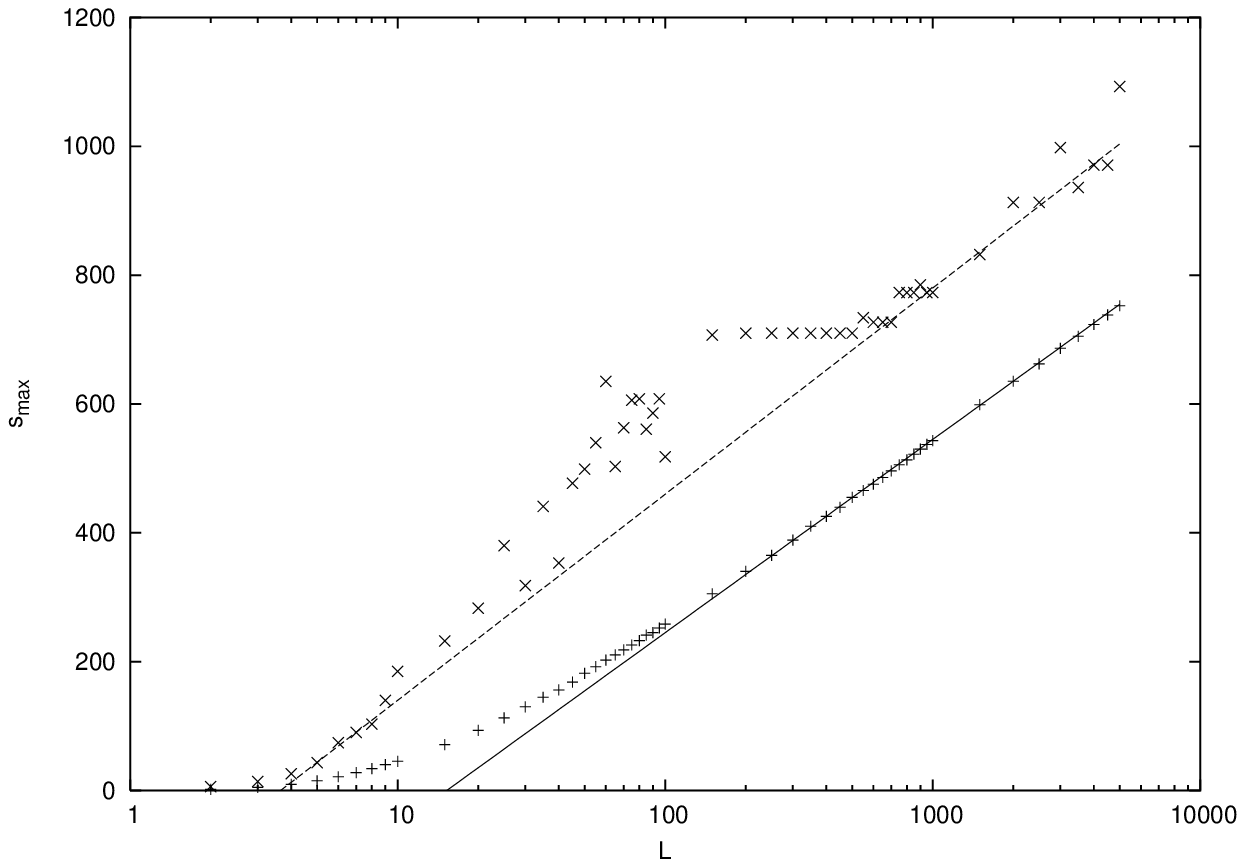}}
\vspace*{8pt}
\caption{Size $s_{\mathrm{max}}$ of the largest cluster, at p=0.25,
 depending on simulated lattice size. $+$ are for $s_{\mathrm{max}}$ averaged
over 1000 runs, $\times$ for the maximal $s_{\mathrm{max}}$ of 1000 runs, which
has stronger fluctuations (as the maximum is determined by single clusters; in 
fact, as no averaging is done, the results for $L$ and $L+1$ are
{\it not} statistically independent).
The solid line corresponds to $300 \log_{10}(L)-355$,
the dashed line to $320 \log_{10}(L)-180$. Thus, for $p<p_c$, the largest
cluster grows $\propto \log(L)$.
\label{f6}}
\end{figure}

For $p<p_c$, we expect the largest cluster (which does not span the whole
lattice) to grow $\propto \log L$, cf.~fig.~\ref{f6}. This is true for $p=0.25$
and $p=0.3$ (not plotted), only slope and intercept are different. From
fig.~\ref{f6} another important effect can be seen: although it seems
natural to determine the largest cluster $s_{\mathrm{max}}$ by taking the
largest cluster of all independent runs, this would mean that no averaging
happens, and thus the values of $s_{\mathrm{max}}$ are not statistically
independent for different $L$: a large cluster in one single run would
dominate the results. Thus, even here averaging is necessary to obtain 
meaningful results.

\section{Summary}

It is possible to modify the Hoshen-Kopelman algorithm in order to obtain
not only results for a $L_{\mathrm{max}}^d$ lattice
in one simulation run, but also results
for intermediate $L$ with $L<L_{\mathrm{max}}$. Compared to the original
HK algorithm, there is a small performance penalty and a higher memory
consumption (a constant factor of $d$ for storing the lattices sites).
The modified algorithm is very useful in order to study the dependance of
percolation properties on the lattice size.

In this paper, the new algorithm was used for site percolation on the
cubic lattice, i.~e.~$d=3$, lattice sizes up to
$5000^3$, and occupation probabilities below, at, and above $p_c$. Results
are presented for cluster size distribution, number of infinite clusters,
and size of the largest cluster.

\section{Outlook}

A natural next step would be to move away from three dimensions and 
to adapt the algorithm to different $d$.
Adapting
it to $d=2$ is easy, while $d>3$ is not easy, but still straightforward.

Parallelization by using domain decomposition is apparently very hard, as
the size of the domains would be changing. This would only be necessary in
order to simulate huge lattices, where the $L$-dependence might not be
very interesting. As several runs always need to be averaged, in order to
obtain statistically independent data, replication is always a viable 
parallelization strategy.

\section*{Acknowledgments}
I would like to thank D.~Stauffer for finally forcing me to publish this 
paper. Furthermore, I would like to thank the Computing Centre of the
University of Cologne for computing time on their HPC-cluster Clio.

\appendix

\end{document}